\documentclass[conference]{IEEEtran}
\IEEEoverridecommandlockouts
\usepackage{cite}
\usepackage{amsmath,amssymb,amsfonts}
\usepackage{algorithm}
\usepackage{algorithmic}
\usepackage{graphicx}
\usepackage{textcomp}
\usepackage{xcolor}
\usepackage{url}
\usepackage{multirow}
\usepackage{flushend}

\def\BibTeX{{\rm B\kern-.05em{\sc i\kern-.025em b}\kern-.08em
   T\kern-.1667em\lower.7ex\hbox{E}\kern-.125emX}}

\begin{document}

\title{Robust and Lossless Fingerprinting of Deep Neural Networks via Pooled Membership Inference}

\author{
Hanzhou Wu\\
Shanghai University, Shanghai 200444, China\\
h.wu.phd@ieee.org\\}

\maketitle

\begin{abstract}
Deep neural networks (DNNs) have already achieved great success in a lot of application areas and brought profound changes to our society. However, it also raises new security problems, among which how to protect the intellectual property (IP) of DNNs against infringement is one of the most important yet very challenging topics. To deal with this problem, recent studies focus on the IP protection of DNNs by applying digital watermarking, which embeds source information and/or authentication data into DNN models by tuning network parameters directly or indirectly. However, tuning network parameters inevitably distorts the DNN and therefore surely impairs the performance of the DNN model on its original task regardless of the degree of the performance degradation. It has motivated the authors in this paper to propose a novel technique called \emph{pooled membership inference (PMI)} so as to protect the IP of the DNN models. The proposed PMI neither alters the network parameters of the given DNN model nor fine-tunes the DNN model with a sequence of carefully crafted trigger samples. Instead, it leaves the original DNN model unchanged, but can determine the ownership of the DNN model by inferring which mini-dataset among multiple mini-datasets was once used to train the target DNN model, which differs from previous arts and has remarkable potential in practice. Experiments also have demonstrated the superiority and applicability of this work.
\end{abstract}


\section{Introduction}
The rapid development of computer hardware and big data technology in the past two decades has promoted deep learning as a popular computing paradigm to achieve great success in a number of application fields such as visual understanding, pattern analysis, natural language processing and bioinformatics. Especially, as one representative architecture of deep learning, deep neural networks (DNNs) are widely deployed in the cloud by many technology companies in order to provide smart and personalized services. It can be foreseen that deep learning will continue making remarkable achievements in different fields. However, creating these state-of-the-art deep models consumes a lot of high-quality data, powerful computing resources and expert knowledge of the architecture design, indicating that as the intellectual property (IP) of the owners, these deep models should be well protected against IP infringement, motivated by which increasing researchers pay attention to the IP protection of DNN models in the past few years.

Mainstream techniques such as \cite{Uchida2017, Adi2018, Wang2020, Zhang:AsiaCCS, Zhao2021ISDFS} suggest digital watermarking for IP protection of the DNN models. The main idea is to embed a secret message revealing the source information or ownership into the DNN model to be protected by directly or indirectly modifying the DNN parameters through an imperceptible way. As a result of embedding a watermark, the performance of the DNN model on its original mission can be well maintained and meanwhile the embedded information can be extracted to identify the copyright of the target DNN.

In terms of watermark embedding, the existing DNN watermarking methods can be divided into two categories. The first category modifies the internal weights \cite{Uchida2017, Wang2020, Li2021} or structures \cite{Zhao2021WIFS} of the DNN to accommodate a watermark. The embedded watermark should be then extracted from the marked weights or structures. Several existing works embed the watermark into the statistical distributions of network parameters, they are still belonging to weight modification. As a result, the ownership should be verified under the white-box scenario, which means that one has to fully (or partly) master the internal details of the target DNN model for watermark extraction. The second category is to embed a secret zero-bit watermark into a given DNN by mixing a set of carefully crafted trigger samples into clean samples for model training \cite{Adi2018, Zhao2021ISDFS, Wang:Sym2022}. As a result, the trained DNN model not only learns the original task of the DNN but also remembers the mapping relationship between the trigger samples and the pre-specified labels. In this way, the ownership is verified by analyzing the consistency between the prediction results of the trigger samples and the labels.

Regardless of the watermarking performance, many existing methods, however, inevitably distort the original DNN model and therefore surely degrade the performance of the host DNN model on its original task regardless of the degradation degree. Moreover, recent studies such as \cite{wnnAttack1, wnnAttack2} have demonstrated that embedding watermark information into DNNs will introduce abnormal statistical characteristics to DNN parameters, which enables the adversary to detect the existence of embedded watermark and even locate the watermark information. Therefore, to deal with this problem, it is desirable to design such an IP protection framework that 1) the functionality of the DNN model would not be impaired unless it was intentionally impaired, 2) the DNN model to be put into use will not expose abnormal statistical characteristics unless it was intentionally modified, and 3) the ownership of the target DNN model can be reliably verified even if it was intentionally attacked by the adversary. Motivated by this insight, a straightforward idea for IP protection is then to keep the original DNN unchanged and develop a robust fingerprinting algorithm for the DNN.

In this paper, we introduce a novel fingerprinting algorithm for DNN models without modifying the DNN models. Unlike previous algorithms, e.g., \cite{Uchida2017, Wang2020, Li2021, Wang:Sym2022, wuTCSVT2021, fragile:paper2, fragile:paper3, fragile:paper1} that either inevitably modify the original DNN model to be protected or are used for DNN integrity verification, the proposed method does not impair the generalization ability of the DNN model at all and can be used for robust verification. The main idea of the proposed method is to collect multiple mini-datasets related to the original task of the model to be protected and determine which mini-dataset among the multiple mini-datasets was used to train the DNN model. We define this technique as \emph{pooled membership inference (PMI)}, which differs from conventional membership inference that determines whether a single data record was part of the model's training dataset or not. In brief summary, the main contributions of this paper include:

\begin{itemize}
	\item Different from many previous watermarking methods that inevitably modify the original DNN model, we keep the DNN model unchanged, which can well preserve the task functionality of the DNN and show high security. That is why we say our method is lossless (to the DNN model).
	\item Different from many previous watermarking methods that use carefully-crafted individual trigger samples to claim the copyright, we use pooled evidence from clean samples to verify the ownership, which is low-cost and efficient.
	\item Different from conventional membership inference inferring whether a single record was belonging to the training set or not, we exploit the distribution relationship between multiple mini-datasets in the embedding space to enhance the statistical evidence of membership inference which is robust for ownership verification. At the same time, there is no need to train shadow models, which indicates that the complexity of the proposed method is very low.
	\item Experimental results have demonstrated that the proposed method not only preserves the performance of the DNN model on its original task, but also enables us to reliably claim the copyright of the target DNN model, which has shown the superiority and applicability of this work.
\end{itemize}

The rest structure of this paper will be organized as follows. In Section II, we introduce the preliminary concepts, followed by the proposed work in Section III. We provide experiments in Section IV. Finally, we conclude this work in Section V.

\section{Preliminary Concepts}
\subsection{Deep Neural Networks}
A deep neural network $\mathcal{M}$ is a biologically-inspired mathematical function that maps an input $\textbf{x}\in \mathcal{X}$ to an output $\textbf{y}\in \mathcal{Y}$, that is, $\textbf{y} = \mathcal{M}(\textbf{x}; \textbf{W})$ always holds, where $\textbf{W}$ is defined as the optimized parameter set of the DNN $\mathcal{M}$. In order to determine $\textbf{W}$, a dataset $D = \{(\textbf{x}_i, \textbf{y}_i)~|i\in [1, |D|]\}$ should be collected to train $\mathcal{M}$ in a way that the optimized $\textbf{W}$ minimizes the total loss between the prediction results and the ground-truths for the samples in $D$. In other words, $\textbf{W}$ is determined by
\begin{equation}
\textbf{W}^* = \underset{\textbf{W}}{\text{arg min}}~\sum_{i=1}^{|D|}L(\mathcal{M}(\textbf{x}_i;\textbf{W}),\textbf{y}_i),
\end{equation}
where $L$ is the loss function measuring the difference between two real vectors. It is noted that the form of $\textbf{y}$ depends on the mission of $\mathcal{M}$, e.g., if $\mathcal{M}$ is limited to image classification, $\textbf{y}$ will be a one-hot vector.

After optimizing the parameter set $\textbf{W}$, we hope the resulting well-trained $\mathcal{M}$ performs very well on unseen data. However, there always exists statistical difference between $D$ and unseen data. From the perspective of practice, blindly optimizing Eq. (1) will cause the over-fitting problem, i.e., the trained model performs pretty good on $D$, but very bad on unseen dataset. To deal with this problem, the dataset $D$ is often divided into two subsets called \emph{training set} and \emph{validation set}. By minimizing the loss on the training set in an iterative way, the validation set enables us to collect the model that has the best generalization ability on unseen data. Also, there are other strategies to select models with good generalization ability, e.g., regularization.

\subsection{DNN Watermarking}
Digital watermarking conceals a signal typically also called watermark within another noise-tolerant signal such as image, video and text. By extracting the hidden watermark from the target watermarked and probably-attacked signal, we are able to identify the ownership of the signal, which promotes digital watermarking to play a quite important role in protecting the intellectual property of digital commercial products. In the past three decades, a number of advanced watermarking algorithms are developed to protect multimedia content. They often model the multimedia data to be marked as a \emph{static signal} which is a sequence of real-valued numbers. Since these cover elements are highly correlated to each other, individual cover elements can be accurately predicted from the local context. As a result, these signals are easy to model, enabling the watermark to be easily inserted into those suitable components of the signals for intellectual property protection without impairing their value.

However, DNN watermarking should take into account the influence on the task functionality of the given DNN model. By reviewing many conventional schemes, DNN watermarking should be evaluated from various aspects at least including:

\begin{itemize}
	\item[-] \emph{Fidelity:} On one hand, \emph{model fidelity} measures the performance degradation of a DNN model on its original task. It requires that the generalization of the DNN model on its original task after watermarking should be kept well. On the other hand, \emph{watermark fidelity} measures the distortion between the watermark extracted from the target model and the original one, which should be as low as possible.
	\item[-] \emph{Imperceptibility:} It should be difficult for the adversary to perceive the existence of the embedded watermark. Otherwise, the adversary may have the chance to locate the watermark and further remove the watermark.
	\item[-] \emph{Payload:} It will be very desirable to embed as many bits as possible into the DNN model. In other words, the size of the watermark is expected to be high.
	\item[-] \emph{Security:} It should be difficult for unauthorized parties to extract, tamper and forge the watermark.
	\item[-] \emph{Robustness:} The watermark embedding operation should be robust against common attacks such as fine-tuning and model compression for reliable ownership verification.
	\item[-] \emph{Complexity:} The computational complexity of marking a DNN model should be as low as possible for better use.
\end{itemize}

Accordingly, considering the fact that a DNN model consists of three modules, i.e., \emph{input}, \emph{internal network} and \emph{output}, the most intuitive strategy to watermark a given DNN is modifying the internal network parameters of the DNN model \cite{Uchida2017, Wang2020, Li2021}. In addition to parameter modification, embedding watermark information into the internal network structure is also desirable \cite{Zhao2021WIFS}. Since a DNN model possesses the ability to accomplish a specific task, we are able to use the functionality of the DNN model for watermark embedding and watermark verification. Along this direction, many methods watermark a DNN based on the mapping relationship between the input and the output \cite{Adi2018, Zhao2021ISDFS, Wang:Sym2022, Zhang:AWEncoder, Liu:Fourier}. Unlike the above methods that focus on robustness, fragile DNN watermarking \cite{fragile:paper1, fragile:paper2, fragile:paper3} has also been studied, which allows for integrity verification of the model.

\subsection{Membership Inference}
Given a trained DNN model and a data record, membership inference \cite{RezaShokri:2017} determines whether the data record was in the training set of the trained DNN model. This can raise privacy risks to individuals. For example, by identifying the fact that a clinical record has been used to train a DNN model associated with a certain disease, membership inference attacks can infer that the owner of the clinical record may have the disease \cite{MI:survey}.

From a technical perspective, membership inference attacks can be performed in either white-box scenario or black-box scenario. In white-box scenario, the attacker is able to collect enough useful information such as the data distribution of the training set and/or the internal details of the target model, to attack the target model. In black-box scenario, the attacker, however, has only black-box access to the target DNN model with limited prior knowledge about the target DNN model.

Through analyzing the similarity between DNN watermarking and membership inference, it is very easy to think of extending membership inference techniques to the IP protection of DNN models. On one hand, the role of the attacker needs to be adjusted. Namely, the attacker in membership inference will become the defender in DNN watermarking. On the other hand, the ownership may be verified by analyzing the internal outputs or the final predictions given a certain number of data records belonging to the training dataset, which will raise two problems. The first one is how to collect the individual records that can be used for robust ownership verification. The second one is how to design the membership inference strategy.

For the first problem, due to the diversity between individual records, it will be not easy to collect ``robust'' data records. For the second problem, it is known that membership inference is effective for over-fitting models, but not good for models with good generalization ability. As a result, the so-called shadow models may be trained in advance to provide assistance, which consumes lots of data and time. To tackle this problem, in this paper, we propose \emph{pooled membership inference (PMI)}, which collects pooled evidence from individual records for inference. Compared with conventional membership inference and DNN watermarking methods, PMI possesses several advantages:

\begin{itemize}
	\item[-] \emph{Availability:} Different from black-box DNN watermarking that needs carefully crafted trigger samples, PMI does not require carefully crafted trigger samples. Instead, any sample related to the task of the DNN may be collected to perform membership inference for IP protection.
	\item[-] \emph{Robustness:} Due to the data diversity, conventional membership inference handling individual records may result in a low inference accuracy. However, PMI focuses on dataset-level membership inference, which can reduce the influence caused by data diversity.
	\item[-] \emph{Security:} PMI does not modify the DNN to be protected, which will never arouse the suspicion from the adversary and therefore demonstrates superior security.
	\item[-] \emph{Complexity:} Different from many membership inference methods that have to train good shadow models, PMI is done by an unsupervised way, whose complexity is low.
\end{itemize}

Based on the aforementioned analysis, we are now ready to introduce the proposed method in the next section.

\begin{figure}[!t]
\centering
\includegraphics[width=\linewidth]{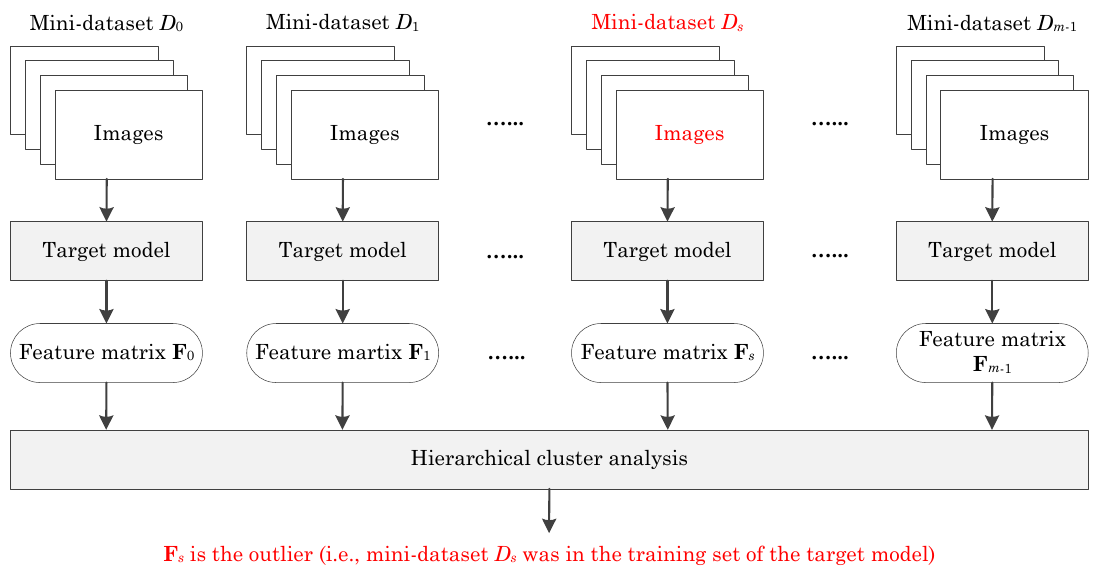}
\caption{General framework for pooled membership inference.}
\end{figure}

\section{Pooled Membership Inference and Its Application to Robust and Lossless Fingerprinting of Deep Neural Networks}
\subsection{Problem Formulation}
Without the loss of generalization, we limit the DNN model $\mathcal{M}$ to be protected to image classification. Therefore, we can redefine $\mathcal{M}$ as a function that maps an input image $\textbf{x}\in \mathcal{X}$ to an integer $y\in \mathcal{Y} = \{0, 1, 2, ..., c-1\}$, where $c \geq 2$ represents the total number of classes. Namely, we have $y = \mathcal{M}(\textbf{x}; \textbf{W})$. Suppose that $\mathcal{M}$ has already well trained, meaning that $\textbf{W}$ is the optimized parameters. Our mission is to infer which one among multiple mini-datasets $\{D_0, D_1, ..., D_{m-1}\}$ was in the training set of $\mathcal{M}$. It is required that one and exactly one mini-dataset was in the training set of $\mathcal{M}$ while the others were not in the training set but related to the original mission of $\mathcal{M}$. It is also assumed that $|D_0| = |D_1| = ... = |D_{m-1}| = n$ and all samples in the mini-datasets have the same ground-truth label.

Mathematically, we can write $D_i$, $1\leq i < m$, as:
\begin{equation}
D_i = \{(\textbf{x}_{i,j}, y_{i,j})|0\leq j< n\},
\end{equation}
where $\textbf{x}_{i,j}$ is the $j$-th sample with ground-truth $y_{i,j}$. And,
\begin{equation}
y_{i,j} = y_{i',j'} = r \in [0,c),~\forall~i\neq i'~\text{or}~j\neq j'.
\end{equation}

Our problem is therefore how to separate $D_s$ from the other mini-datasets to support the fact that the mini-dataset $D_s$ was in the training set of $\mathcal{M}$ while the others were not.

\subsection{Pooled Membership Inference}
We would like to clarify that the PMI is not to infer whether a mini-dataset was belonging to the training set of a trained model only using the mini-dataset itself. Instead, PMI aims at identifying the required mini-dataset among multiple ones.

Fig. 1 shows the general framework of PMI. We first reduce each mini-dataset $D_i, i\in [0, m),$ to a 2-D feature matrix $\textbf{F}_i$. The well trained model $\mathcal{M}$ offers the off-the-shelf solution. In detail, for each sample $\textbf{x}_{i,j}\in D_i$, we feed it to $\mathcal{M}$ and get the vector outputted by the last layer of $\mathcal{M}$ without softmax as the low-dimensional representation of $\textbf{x}_{i,j}$, expressed as $\textbf{f}_{i,j} = \mathcal{M}_\mathrm{last}(\textbf{x}_{i,j}; \textbf{W}) \in \mathbb{R}^c$. Then, $\textbf{F}_i$ is determined by collecting all the low-dimensional representations to form a feature matrix:
\begin{equation}
\textbf{F}_i =
\begin{bmatrix}
\textbf{f}_{i,0}, \textbf{f}_{i,1}, \cdots, \textbf{f}_{i,n-1}
\end{bmatrix}^\text{T},
\end{equation}
whose dimension is $n\times c$. Accordingly, we are able to collect a total of $m$ feature matrices $\textbf{F}_0$, $\textbf{F}_1$, ..., $\textbf{F}_{m-1}$. We argue that if a mini-dataset was once used to train $\mathcal{M}$, the distribution of its feature matrix in somehow feature space will be significantly different from other mini-datasets not appearing in the training set. As a result, the feature point far away from the other points can be judged as an outlier, i.e., the corresponding mini-dataset should be in the training set. It can be solved by applying either \emph{clustering} or \emph{outlier detection} \cite{Ker:TIFS}. We here apply clustering.

Before clustering, we have to normalize the feature matrices and determine the distance between different feature matrices. By feature normalization, each component of the features has zero mean and unit variance, i.e.,
\begin{equation}
\frac{1}{nm}\sum_{i=0}^{m-1}\sum_{j=0}^{n-1}\textbf{F}_{i,j,k} = 0,~\forall ~0\leq k < c
\end{equation}
and
\begin{equation}
\frac{1}{nm}\sum_{i=0}^{m-1}\sum_{j=0}^{n-1}\textbf{F}_{i,j,k}^2 = 1,~\forall ~0\leq k < c
\end{equation}
where $\textbf{F}_{i,j,k}$ is defined as the $k$-th element of $\textbf{f}_{i,j}$. Normalization enables the following distance measure to be more meaningful and not significantly affected by noisy components \cite{Wu:chapter}. Maximum mean discrepancy (MMD) \cite{MMDpaper, MMDsurvey} is effective for measuring the distance between two sets of feature vectors. In general, given two sets $X=\{\textbf{x}_i\}_{i=1}^{|X|}$ and $Y=\{\textbf{y}_i\}_{i=1}^{|Y|}$ which are i.i.d. drawn from $p(\textbf{x})$ and $q(\textbf{y})$ on $\mathbb{R}^d$, let $\mathcal{F}$ be a class of functions $f:\mathbb{R}^d\mapsto \mathbb{R}$, the MMD and its estimate are \cite{Wu:chapter}:
\begin{equation}
\text{MMD}[\mathcal{F},p,q] = \underset{f\in \mathcal{F}}{\text{sup}}~\mathbb{E}_{\textbf{x}\sim p(\textbf{x})}f(\textbf{x})-\mathbb{E}_{\textbf{y}\sim q(\textbf{y})}f(\textbf{y}),
\end{equation}
\begin{equation}
\text{MMD}[\mathcal{F},X,Y] = \underset{f\in \mathcal{F}}{\text{sup}}~\frac{1}{|X|}\sum_{\textbf{x}\in X}f(\textbf{x})-\frac{1}{|Y|}\sum_{\textbf{y}\in Y}f(\textbf{y}).
\end{equation}
$\mathcal{F}$ can be selected as a unit ball in a universal RKHS $\mathcal{H}$ defined on the compact metric space $\mathbb{R}^d$ with kernel $k(\cdot,\cdot)$ and feature mapping $\phi(\cdot)$. It has been proven that
\begin{equation}
\text{MMD}^2[\mathcal{F},p,q] = \left \|\mathbb{E}_{\textbf{x}\sim p(\textbf{x})}\phi(\textbf{x})-\mathbb{E}_{\textbf{y}\sim q(\textbf{y})}\phi(\textbf{y}) \right \|_{\mathcal{H}}^2.
\end{equation}
An \emph{unbiased} estimate of MMD is:
\begin{equation}
\text{MMD}[\mathcal{F},X,Y] = \left (\frac{1}{|X|^2-|X|}\sum_{i\neq j}h[i,j]\right )^{1/2},
\end{equation}
where $|X| = |Y|$ and
\begin{equation}
h[i,j] = k(\textbf{x}_i,\textbf{x}_j) + k(\textbf{y}_i,\textbf{y}_j) - k(\textbf{x}_i,\textbf{y}_j) - k(\textbf{x}_j,\textbf{y}_i).
\end{equation}
For any two sets, the unbiased estimate of MMD can be used to measure their distance, for which a kernel function $k(\cdot,\cdot)$ is required. In this paper, we use the unbiased estimate of MMD to determine the distance between two feature sets $\textbf{F}_i$ and $\textbf{F}_j$. And, the dot product, i.e., $k(\textbf{x},\textbf{y}) = \textbf{x}\cdot\textbf{y}$ is used as the kernel.

\begin{algorithm}[!t]
 \caption{Pseudo-code of Agglomerative Clustering}
 \begin{algorithmic}[1]
	\STATE Collect objects $\{c_i\}_{i=1}^{n_c}$, and set distance measure $D(\cdot,\cdot)$
	\STATE $\mathcal{C} \leftarrow \{\{c_1\}, \{c_2\}, ..., \{c_{n_c}\}\}$ \hfill $\triangleright$ Each as its own cluster
    \STATE $\mathcal{T} \leftarrow \mathcal{C}$ \hfill $\triangleright$ Initialize the tree
	\WHILE {$|\mathcal{C}|>1$}
	   \STATE $\mathcal{G}_1^*,\mathcal{G}_2^*\leftarrow \underset{\mathcal{G}_1,\mathcal{G}_2\in \mathcal{C}}{\text{arg min}}~D(\mathcal{G}_1,\mathcal{G}_2)$ \hfill $\triangleright$ Minimize $D$
        \STATE $\mathcal{C}\leftarrow\mathcal{C}\setminus\{\mathcal{G}_1^*,\mathcal{G}_2^*\}$ \hfill $\triangleright$ Remove both from the active set
        \STATE $\mathcal{C}\leftarrow\mathcal{C}\cup\{\mathcal{G}_1^*\cup\mathcal{G}_2^*\}$ \hfill $\triangleright$ Add union to the active set
        \STATE $\mathcal{T}\leftarrow\mathcal{T}\cup\{\mathcal{G}_1^*\cup\mathcal{G}_2^*\}$ \hfill $\triangleright$ Add union to the tree
	\ENDWHILE
	\RETURN $\mathcal{T}$
 \end{algorithmic}
\end{algorithm}

After computing all the distances between feature matrices, we are to perform hierarchical cluster analysis \cite{clustering:book} to identify the required feature matrix. As a kind of hierarchical analysis approach, \emph{agglomerative clustering} is empirically used in this paper. It is a bottom-up technique that each feature point starts in its own cluster and pairs of the clusters are merged as one moves up the hierarchy. Pseudo-code is given in Algorithm 1, from which we can infer that in each iteration step the nearest two clusters are merged and it iterates until there is only one cluster. A key problem is how to define the distance measure $D(\cdot,\cdot)$ between two clusters. In this paper, we use the single linkage by default due to its simplicity and efficiency, i.e.,
\begin{equation}
D(X, Y) = \underset{x\in X, y\in Y}{\text{min}}~d(x,y),
\end{equation}
where $d(x, y)$ denotes the distance between two objects $x$ and $y$, and $D(X, Y)$ for the distance between two clusters $X$ and $Y$. When to apply Eq. (12) to two feature matrices $\textbf{F}_i$ and $\textbf{F}_j$, $d(\textbf{F}_i, \textbf{F}_j)$ is equivalent to the corresponding MMD distance.

By applying agglomerative clustering, two clusters $\mathcal{G}_1^*$ and $\mathcal{G}_2^*$ at the ``final'' stage of merging can be obtained. The objects belonging to the cluster with a smaller size are considered as the abnormal objects. For example, assuming that $|\mathcal{G}_1^*|\leq |\mathcal{G}_2^*|$, any mini-dataset in $\mathcal{G}_1^*$ may be judged as part of the training set of $\mathcal{M}$. Since we demand that only one mini-dataset was in the training set, it is always expected that $|\mathcal{G}_1^*| \equiv 1$. However, if $|\mathcal{G}_1^*| > 1$, a mini-dataset can be randomly selected out from $\mathcal{G}_1^*$ and judged as part of the training set. Thus, PMI is completed.

\subsection{Robust and Lossless Fingerprinting of DNNs}
Based on the proposed PMI, we are now ready to implement a robust and lossless fingerprinting system for DNN models. Assuming that we have collected two datasets $P$ and $Q$ related to the task of $\mathcal{M}$. All samples in $P\cup Q$ are belonging to the same class $r\in [0, c)$. The only one difference between $P$ and $Q$ is that $P$ was used to train $\mathcal{M}$ but $Q$ was not. We conduct an experiment $t$ times. In each time, $n$ different samples in $P$ are randomly selected out to form a mini-dataset. And, $n(m-1)$ different samples in $Q$ are randomly selected out to form $m-1$ disjoint mini-datasets, each of which has exactly $n$ samples\footnote{Each mini-dataset is associated with a randomly generated index in $[0, m)$.}. Therefore, it is necessary that $|P|\geq n$ and $|Q|\geq n(m-1)$. For each experiment, if the mini-dataset corresponding to $P$ is identified, we say the experiment is successful. The percentage of successful experiments can be calculated, and used to verify the ownership. Clearly, let $\mathrm{acc}_r$ be the percentage of successful experiments. The ownership can be successfully identified if
\begin{equation}
\mathrm{acc}_r - \frac{1}{m} \geq \rho,
\end{equation}
where $1/m$ represents the percentage of ``random guess'' and $\rho\in (0, 1-1/m]$ is a pre-determined threshold. In other words, the larger $\mathrm{acc}_r - 1/m$, the better the fingerprinting behavior.

\section{Experimental Results and Analysis}
In this section, we will conduct our experiments and analysis to evaluate the performance of the proposed method.

\subsection{Datasets and Models}
We use two most popular datasets MNIST\footnote{Online available: \url{http://yann.lecun.com/exdb/mnist/}} and CIFAR-10\footnote{Online available: \url{https://www.cs.toronto.edu/~kriz/cifar.html}} for simulation. The former dataset consists of a total of 70,000 grayscale images each with a size of $28\times 28\times 1$ in ten classes. The latter dataset contains a total of 60,000 color images each with a size of $32\times 32\times 3$ in ten classes. We train two popular models MLP and LeNet-5 \cite{LeNet} on MNIST, and train another two models VGG-19 \cite{VGG} and ResNet-34 \cite{ResNet} on CIFAR-10. MLP consists of an input layer, a hidden layer with 64 nodes and an output layer. All these models use ReLU \cite{ReLU} as the activation function for internal layers. Besides, we use Adam \cite{Adam} for parameter optimization during model training and the open framework PyTorch\footnote{Online available: \url{https://pytorch.org/}} for simulation.

For model training, we divide each dataset into two disjoint subsets called \emph{training set} and \emph{validation set}. The training set is used to optimize the parameters of the host DNN model and the validation set is used to help us to select a trained model with the best generalization ability. It is noted that we did not use \emph{testing set}, which helps us to evaluate the generalization ability of the model. The reason is that the proposed PMI does not modify the model to be protected at all. As a result, the generalization ability of the target model will not be impaired at all unless the model was modified by the adversary himself. Therefore, we will not report the generalization performance of the trained DNN model on unseen data. In our experiments, for the MNIST dataset, the ratio between the size of the training set and the size of the validation set is 50,000:20,000. And for CIFAR-10, the ratio is empirically set to 45,000:15,000.

In addition, in experiments, the batch size is set to 64. The learning rate is initialized as 0.001. The total number of epochs is fixed as 300. We use $t$ = 100. Though all these parameters can be fine-tuned, our experimental results have confirmed that they already result in superior fingerprinting performance.

\begin{table}[!t]
\renewcommand{\arraystretch}{1}
\centering
\caption{The fingerprinting performance due to different parameters evaluated on the MNIST dataset.}
\begin{tabular}{c|c|c|c|c|c|c}
\hline\hline
DNN Model & $m$ & $n$& $r_\text{opt}$ & $\mathrm{acc}_{r_\text{opt}}$ & $\overline{\mathrm{acc}_r}$ & $1/m$\\
\hline
\multicolumn{1}{c|}{\multirow{9}{*}{MLP}}
& \multicolumn{1}{c|}{\multirow{3}{*}{3}} & 100 & 6 & 0.50 & 0.392 & \multicolumn{1}{c}{\multirow{3}{*}{0.33}}\\
&  & 200 & 6 & 0.58 & 0.409 \\
&  & 300 & 6 & 0.73 & 0.471 \\
\cline{2-7}
& \multicolumn{1}{c|}{\multirow{3}{*}{4}} & 100 & 3 & 0.48 & 0.305 & \multicolumn{1}{c}{\multirow{3}{*}{0.25}}\\
&  & 200 & 6 & 0.46 & 0.346 \\
&  & 300 & 6 & 0.66 & 0.407 \\
\cline{2-7}
& \multicolumn{1}{c|}{\multirow{3}{*}{5}} & 100 & 3 & 0.42 & 0.244 & \multicolumn{1}{c}{\multirow{3}{*}{0.20}}\\
&  & 200 & 6 & 0.47 & 0.286 \\
&  & 300 & 6 & 0.72 & 0.367 \\
\hline
\multicolumn{1}{c|}{\multirow{9}{*}{LeNet-5}} &
\multicolumn{1}{c|}{\multirow{3}{*}{3}} & 100 & 0 & 0.47 & 0.363 &  \multicolumn{1}{c}{\multirow{3}{*}{0.33}}\\
&  & 200 & 6 & 0.55 & 0.391 \\
&  & 300 & 6 & 0.63 & 0.434 \\
\cline{2-7}
& \multicolumn{1}{c|}{\multirow{3}{*}{4}} & 100 & 3 & 0.35 & 0.273 &  \multicolumn{1}{c}{\multirow{3}{*}{0.25}} \\
&  & 200 & 0 & 0.51 & 0.366 \\
&  & 300 & 6 & 0.66 & 0.394 \\
\cline{2-7}
& \multicolumn{1}{c|}{\multirow{3}{*}{5}} & 100 & 6 & 0.36 & 0.243 & \multicolumn{1}{c}{\multirow{3}{*}{0.20}} \\
&  & 200 & 6 & 0.45 & 0.285 \\
&  & 300 & 6 & 0.51 & 0.314 \\
\hline\hline
\end{tabular}
\end{table}

\begin{table}[!t]
\renewcommand{\arraystretch}{1}
\centering
\caption{The fingerprinting performance due to different parameters evaluated on the CIFAR-10 dataset.}
\begin{tabular}{c|c|c|c|c|c|c}
\hline\hline
DNN Model & $m$ & $n$& $r_\text{opt}$ & $\mathrm{acc}_{r_\text{opt}}$ & $\overline{\mathrm{acc}_r}$ & $1/m$\\
\hline
\multicolumn{1}{c|}{\multirow{9}{*}{VGG-19}}
& \multicolumn{1}{c|}{\multirow{3}{*}{3}} & 100 & 5 & 0.93 & 0.718 & \multicolumn{1}{c}{\multirow{3}{*}{0.33}}\\
&  & 200 & 4 & 0.99 & 0.861 \\
&  & 300 & 5 & 0.99 & 0.928 \\
\cline{2-7}
& \multicolumn{1}{c|}{\multirow{3}{*}{4}} & 100 & 2 & 0.90 & 0.689 & \multicolumn{1}{c}{\multirow{3}{*}{0.25}}\\
&  & 200 & 4 & 0.99 & 0.848 \\
&  & 300 & 6 & 1.00 & 0.922 \\
\cline{2-7}
& \multicolumn{1}{c|}{\multirow{3}{*}{5}} & 100 & 5 & 0.90 & 0.708 & \multicolumn{1}{c}{\multirow{3}{*}{0.20}}\\
&  & 200 & 4 & 0.99 & 0.867 \\
&  & 300 & 5 & 1.00 & 0.931 \\
\hline
\multicolumn{1}{c|}{\multirow{9}{*}{ResNet-34}} &
\multicolumn{1}{c|}{\multirow{3}{*}{3}} & 100 & 4 & 1.00 & 0.899 &  \multicolumn{1}{c}{\multirow{3}{*}{0.33}}\\
&  & 200 & 6 & 1.00 & 0.987 \\
&  & 300 & 8 & 1.00 & 0.998 \\
\cline{2-7}
& \multicolumn{1}{c|}{\multirow{3}{*}{4}} & 100 & 5 & 1.00 & 0.913 &  \multicolumn{1}{c}{\multirow{3}{*}{0.25}} \\
&  & 200 & 7 & 1.00 & 0.991 \\
&  & 300 & 7 & 1.00 & 0.995 \\
\cline{2-7}
& \multicolumn{1}{c|}{\multirow{3}{*}{5}} & 100 & 3 & 1.00 & 0.920 & \multicolumn{1}{c}{\multirow{3}{*}{0.20}} \\
&  & 200 & 6 & 1.00 & 0.994 \\
&  & 300 & 8 & 1.00 & 0.999 \\
\hline\hline
\end{tabular}
\end{table}

\begin{table}[!t]
\renewcommand{\arraystretch}{1}
\centering
\caption{The fingerprinting performance after model fine-tuning evaluated on the MNIST dataset.}
\begin{tabular}{c|c|c|c|c|c|c}
\hline\hline
DNN Model & $m$ & $n$& $r_\text{opt}$ & $\mathrm{acc}_{r_\text{opt}}$ & $\overline{\mathrm{acc}_r}$ & $1/m$\\
\hline
\multicolumn{1}{c|}{\multirow{9}{*}{MLP}}
& \multicolumn{1}{c|}{\multirow{3}{*}{3}} & 100 & 6 & 0.54 & 0.398 & \multicolumn{1}{c}{\multirow{3}{*}{0.33}}\\
&  & 200 & 6 & 0.57 & 0.401 \\
&  & 300 & 6 & 0.70 & 0.437 \\
\cline{2-7}
& \multicolumn{1}{c|}{\multirow{3}{*}{4}} & 100 & 6 & 0.47 & 0.322 & \multicolumn{1}{c}{\multirow{3}{*}{0.25}}\\
&  & 200 & 3 & 0.49 & 0.340 \\
&  & 300 & 6 & 0.67 & 0.385 \\
\cline{2-7}
& \multicolumn{1}{c|}{\multirow{3}{*}{5}} & 100 & 6 & 0.45 & 0.263 & \multicolumn{1}{c}{\multirow{3}{*}{0.20}}\\
&  & 200 & 3 & 0.46 & 0.298 \\
&  & 300 & 3 & 0.66 & 0.377 \\
\hline
\multicolumn{1}{c|}{\multirow{9}{*}{LeNet-5}} &
\multicolumn{1}{c|}{\multirow{3}{*}{3}} & 100 & 2 & 0.50 & 0.371 &  \multicolumn{1}{c}{\multirow{3}{*}{0.33}}\\
&  & 200 & 2 & 0.57 & 0.398 \\
&  & 300 & 6 & 0.67 & 0.440 \\
\cline{2-7}
& \multicolumn{1}{c|}{\multirow{3}{*}{4}} & 100 & 6 & 0.45 & 0.305 &  \multicolumn{1}{c}{\multirow{3}{*}{0.25}} \\
&  & 200 & 6 & 0.59 & 0.341 \\
&  & 300 & 6 & 0.65 & 0.369 \\
\cline{2-7}
& \multicolumn{1}{c|}{\multirow{3}{*}{5}} & 100 & 6 & 0.34 & 0.230 & \multicolumn{1}{c}{\multirow{3}{*}{0.20}} \\
&  & 200 & 6 & 0.44 & 0.317 \\
&  & 300 & 6 & 0.63 & 0.367 \\
\hline\hline
\end{tabular}
\end{table}

\begin{table}[!t]
\renewcommand{\arraystretch}{1}
\centering
\caption{The fingerprinting performance after model fine-tuning evaluated on the CIFAR-10 dataset.}
\begin{tabular}{c|c|c|c|c|c|c}
\hline\hline
DNN Model & $m$ & $n$& $r_\text{opt}$ & $\mathrm{acc}_{r_\text{opt}}$ & $\overline{\mathrm{acc}_r}$ & $1/m$\\
\hline
\multicolumn{1}{c|}{\multirow{9}{*}{VGG-19}}
& \multicolumn{1}{c|}{\multirow{3}{*}{3}} & 100 & 2 & 0.92 & 0.705 & \multicolumn{1}{c}{\multirow{3}{*}{0.33}}\\
&  & 200 & 2 & 0.98 & 0.862 \\
&  & 300 & 5 & 1.00 & 0.927 \\
\cline{2-7}
& \multicolumn{1}{c|}{\multirow{3}{*}{4}} & 100 & 2 & 0.95 & 0.681 & \multicolumn{1}{c}{\multirow{3}{*}{0.25}}\\
&  & 200 & 5 & 0.99 & 0.849 \\
&  & 300 & 5 & 1.00 & 0.919 \\
\cline{2-7}
& \multicolumn{1}{c|}{\multirow{3}{*}{5}} & 100 & 5 & 0.92 & 0.712 & \multicolumn{1}{c}{\multirow{3}{*}{0.20}}\\
&  & 200 & 5 & 0.98 & 0.868 \\
&  & 300 & 4 & 1.00 & 0.937 \\
\hline
\multicolumn{1}{c|}{\multirow{9}{*}{ResNet-34}} &
\multicolumn{1}{c|}{\multirow{3}{*}{3}} & 100 & 5 & 1.00 & 0.918 &  \multicolumn{1}{c}{\multirow{3}{*}{0.33}}\\
&  & 200 & 6 & 1.00 & 0.979 \\
&  & 300 & 7 & 1.00 & 0.994 \\
\cline{2-7}
& \multicolumn{1}{c|}{\multirow{3}{*}{4}} & 100 & 3 & 1.00 & 0.908 &  \multicolumn{1}{c}{\multirow{3}{*}{0.25}} \\
&  & 200 & 7 & 1.00 & 0.989 \\
&  & 300 & 8 & 1.00 & 0.995 \\
\cline{2-7}
& \multicolumn{1}{c|}{\multirow{3}{*}{5}} & 100 & 5 & 1.00 & 0.918 & \multicolumn{1}{c}{\multirow{3}{*}{0.20}} \\
&  & 200 & 6 & 1.00 & 0.987 \\
&  & 300 & 0 & 1.00 & 0.999 \\
\hline\hline
\end{tabular}
\end{table}

\subsection{Effectiveness}
As mentioned in Section III-C, we have to specify the value of $r$. Meanwhile, two datasets $P$ and $Q$ should be collected. To this purpose, taking MNIST for explanation, when $r$ is given, we could use the samples with a ground-truth $r$ in the training set to constitute $P$ and the samples with a ground-truth $r$ in the validation set to constitute $Q$. Accordingly, we are able to verify the ownership (fingerprint) of the target model. Because there are different $r$ for choice, we will test all possible values of $r$ so that the optimal solution $r_\text{opt}$ can be used, i.e.,
\begin{equation}
r_\text{opt} = \arg\max_r~\left(\mathrm{acc}_r - 1/m\right).
\end{equation}

Obviously, we expect that $r_\text{opt}$ is as high as possible. From a statistical significance point of view, PMI is efficient if
\begin{equation}
\overline{\mathrm{acc}_r} = \sum_{i=0}^{c-1}\frac{\mathrm{acc}_i}{c} \geq \frac{1}{m} + \rho.
\end{equation}

Table I and Table II provide the experimental results of fingerprinting evaluated on MNIST and CIFAR-10, respectively. Several conclusions can be figured out. First of all, in all cases shown in the two Tables, $\mathrm{acc}_r$ is significantly higher than $1/m$, indicating that the proposed PMI is effective for fingerprinting. Second, different parameters result in different performance of fingerprinting. Specifically, for a fixed $m$, $\mathrm{acc}_r$ increases as $n$ increases, which is due to the reason that a larger $n$ enhances the feature difference between different clusters, allowing for better clustering performance. On the other hand, for a fixed $n$, $\mathrm{acc}_r$ tends to decline when $m$ increases which is reasonable since more feature matrices increase the difficulty of successful clustering. Third, $\overline{\mathrm{acc}_r}$  is reasonably lower than $\mathrm{acc}_r$ but still significantly higher than $1/m$, especially for CIFAR-10. Here, the reason why $\mathrm{acc}_r$ and $\overline{\mathrm{acc}_r}$ for CIFAR-10 are significantly higher than MNIST can be explained as follows. That is, the rich semantic information of CIFAR-10 enables the extracted high-level features to be more discriminative than MNIST. It then benefits the subsequent clustering process a lot.

\begin{figure*}
\centering
\includegraphics[width=\linewidth]{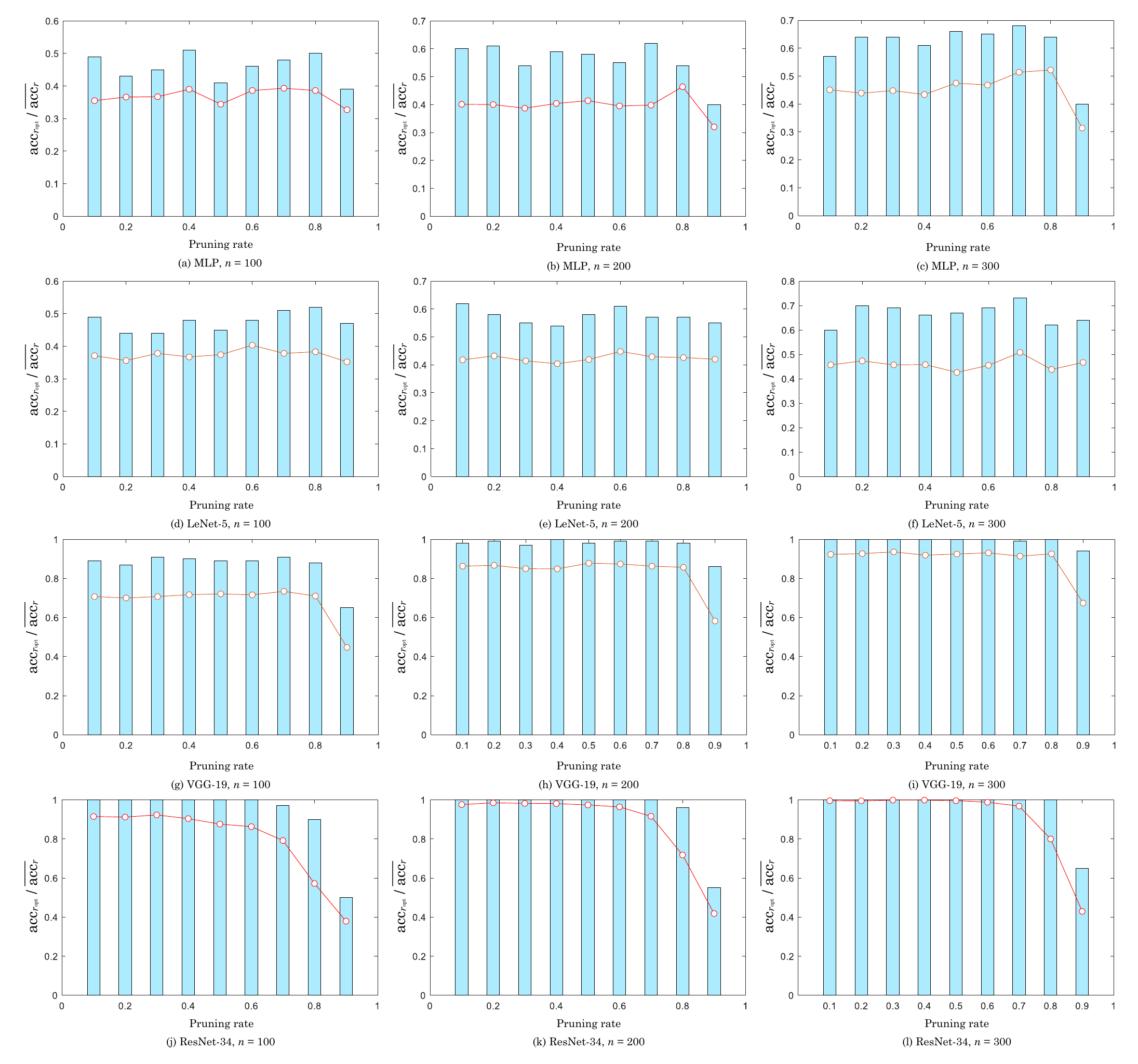}
\caption{The fingerprinting performance after model pruning in case $m = 3$. The column chart corresponds to $\mathrm{acc}_{r_\text{opt}}$, and the line chart corresponds to $\overline{\mathrm{acc}_r}$.}
\end{figure*}

\subsection{Robustness}
In practice, the trained model may be fine-tuned or pruned to achieve better image classification performance or lightweight model deployment. To this end, we evaluate the fingerprinting performance after model fine-tuning and pruning. This enables us to quantify the robustness of the proposed PMI technique. Model fine-tuning aims at adjusting the parameters of a trained model to fit with certain observations, whereas model pruning reduces the size of a model for lightweight deployment.

In experiments, we randomly choose 20\% samples from the validation set for model fine-tuning. Therefore, the set $Q$ needs to be updated, but $P$ should be unchanged. For model pruning, we remove a certain number of internal network weights with the lowest $\ell_1$ norm according to a pruning rate. It is interesting that it is not necessary to use the value of $r_\text{opt}$ determined in Table I and Table II for fingerprinting. In other words, for the fine-tuned or pruned model, we can still enumerate $r$ in range $[0, c)$ and find $r_\text{opt}$ whose value may be different from the one used for the original model. The corresponding $\mathrm{acc}_{r_\text{opt}}$ can be used to verify the ownership of the target DNN model.

Table III and Table IV provide the experimental results after model fine-tuning, from which we can infer that fine-tuning a trained model will not impair the fingerprinting performance. The reason is that same-source fine-tuning enables the model to better fit the training set, accordingly keeping or enhancing the distribution difference between normal feature points and abnormal feature points. In Table IV, there may be multiple $r$ having the largest $\mathrm{acc}_{r}$. Any one of them can be used as $r_\text{opt}$.

The experimental results provided in the previous subsection have indicated that a smaller $m$ is more suitable for fingerprinting of DNNs as $n$ is specified. Besides, a smaller $m$ implies that the computational cost for clustering is lower. Therefore, from the viewpoint of applications, it is better to use a smaller $m$. Based on this perspective, to evaluate the fingerprinting performance against model pruning, we set $m = 3$ in simulation experiments. Figure 2 has demonstrated the fingerprinting performance after model pruning by applying different pruning rates. It can be inferred from Figure 2 that, as the pruning rate increases, the fingerprinting performance tends to decline in most cases, which is reasonable due to the removal of some important network weights. However, it can be inferred that both $\mathrm{acc}_{r_\text{opt}}$ and $\overline{\mathrm{acc}_r}$ are significantly higher than $1/3$ (which corresponds to ``random guess'') in most cases, indicating that our work has strong ability to resist model pruning.

\section{Conclusion}
In this paper, we propose a new technical framework for IP protection of DNN models. The proposed method keeps the original DNN model to be protected unchanged and verify the ownership by inferring whether a mini-dataset among multiple mini-datasets was in the training set of the target DNN model or not, which is completed by mapping each mini-dataset into a feature point and find the point far away from other points as the outlier that corresponds to the required mini-dataset. Our experimental results show that the proposed method can verify the ownership (fingerprint) of the target model effectively and is robust to common attacks including fine-tuning and pruning. We will improve the proposed work to resist more attacks and apply the proposed technique for fingerprinting of datasets.

\section*{Acknowledgement}
This work was financially supported in part by the Opening Project of Guangdong Province Key Laboratory of Information Security Technology under grant number 2020B1212060078, the National Natural Science Foundation of China under Grant No. 61902235, the Shanghai ``Chen Guang'' project supported by Shanghai Municipal Education Commission and Shanghai Education Development Foundation, and the CCF-Tencent Open Research Fund.

\end{document}